\documentclass[aps,prl,twocolumn,showpacs,superscriptaddress]{revtex4}
\usepackage{graphicx}

\begin{document}

\title{Evolution of band structure from optimally doped to heavily overdoped Co-substituted NaFeAs}

\author{S. T. Cui}
\affiliation{National Synchrotron Radiation Laboratory, University of Science and Technology of China, Hefei, Anhui 230029, P. R. China}

\author{S. Y. Zhu}
\affiliation{National Synchrotron Radiation Laboratory, University of Science and Technology of China, Hefei, Anhui 230029, P. R. China}

\author{A. F. Wang}
\affiliation{Hefei National Laboratory for Physical Sciences at Microscale and Department of Physics, University of Science and Technology of China, Hefei, Anhui 230026, P. R. China}

\author{S. Kong}
\affiliation{National Synchrotron Radiation Laboratory, University of Science and Technology of China, Hefei, Anhui 230029, P. R. China}

\author{S. L. Ju}
\affiliation{National Synchrotron Radiation Laboratory, University of Science and Technology of China, Hefei, Anhui 230029, P. R. China}

\author{X. G. Luo}
\affiliation{Hefei National Laboratory for Physical Sciences at Microscale and Department of Physics, University of Science and Technology of China, Hefei, Anhui 230026, P. R. China}

\author{X. H. Chen}
\affiliation{Hefei National Laboratory for Physical Sciences at Microscale and Department of Physics, University of Science and Technology of China, Hefei, Anhui 230026, P. R. China}

\author{G. B. Zhang}
\affiliation{National Synchrotron Radiation Laboratory, University of Science and Technology of China, Hefei, Anhui 230029, P. R. China}

\author{Z. Sun}
\email{zsun@ustc.edu.cn}
\affiliation{National Synchrotron Radiation Laboratory, University of Science and Technology of China, Hefei, Anhui 230029, P. R. China}

\begin{abstract}
Using angle-resolved photoemission spectroscopy, we studied the electronic structure of NaFe$_{1-x}$Co$_x$As from an optimally doped superconducting compound ($x=0.028$) to a heavily overdoped non-superconducting one ($x=0.109$). Similar to the case of ``122" type iron pnictides, our data suggest that Co dopant in NaFe$_{1-x}$Co$_x$As supplies extra charge carriers and shifts the Fermi level accordingly. In the $x=0.109$ compound, the hole-like bands around the zone center $\Gamma$ move to deeper binding energies and an electron pocket appears instead. The overall band renormalization remains basically the same throughout the doping range we studied, suggesting that the local magnetic/electronic correlations are not affected by carrier doping. We speculate that a balance between itinerant properties of mobile carriers and local interactions may play an important role for the superconductivity.

\end{abstract}


\pacs{74.25.Jb, 71.20.-b, 74.70.Xa, 79.60.-i}

\maketitle

Chemical substitutions in novel materials have been widely used to control the Fermi energy, change electron correlations, scatter quasiparticle excitations, apply chemical pressure, and so on, which often result in intriguing properties and complex phase diagrams. For instance, iron-pnictide materials evolve from a magnetically ordered phase to a high temperature superconducting state and enter into a normal metallic phase with various dopants \cite{Kamihara,xhchen,Rotter,Canfield,Shuai,Schnelle}. Direct measurements of electronic structures by angle-resolved photoemission spectroscopy (ARPES) have shown different effects of chemical substitutions $-$ Co dopants induce a Lifshitz transition \cite{LiuNat} and change the chemical potential by providing extra \emph{d} electrons \cite{LiuPRB}, Ru substitution leads to magnetic dilution with evident variation of the band structure at high doping levels \cite{Dhaka,Xu}, and isoelectronic P doping introduces chemical pressure and brings on a strong band renormalization \cite{ZhangNat,Yoshida,Thirupathaiah}. Despite the manifold effects of chemical substitutions, the superconductivity in iron pnictides arises with the destruction of long-range antiferromagnetic ordering in parent compounds and fades away with further chemical substitutions. Therefore, different dopants can emphasize or induce distinct microscopic interactions, helping to distinguish individual actions of various degrees of freedom and their relevancies to superconductivity.

Unlike Ru and P, the primary role of Co substitution is to increase electron carriers without inducing notable changes of local electronic correlations. In the intensively studied ``122" type iron pnictides, ARPES measurements of band structures explicitly show that the Co dopant changes the chemical potential and Fermi surface topology with a rigid-band-like carrier doping behavior \cite{LiuPRB,Neupane,Ideta}. To gain more knowledge about the effects of carrier doping on superconductivity in iron pnictides, investigations should be extended to other families of FeAs compounds, for instance ``111" type, which can provide important insight into the microscopic interactions in these materials \cite{Parker,zhliu,Wright,syZhou,Wang,xZhou}. In particular, Co doping provides a good opportunity to  investigate how the variation of carrier concentrations change the itinerant properties of electrons, the  local magnetic and/or electronic correlations, and the subtle balance between them, which will reveal underlying interactions that are related to the superconductivity.

Here we report ARPES studies of Co-substituted ``111" type NaFeAs to address the impacts of carrier doping on band structures and microscopic interactions. The absence of surface reconstruction or charge redistribution makes NaFe$_{1-x}$Co$_x$As an ideal system for ARPES to investigate the electronic structure of iron-pnictides materials. We focus on the doping levels ranging from the optimal doping ($x=0.028$) to the heavily overdoped non-superconducting regime ($x=0.109$), where the primary change in electronic properties is induced by the carrier doping and influences of the structural/magnetic transitions are eliminated (see Fig. 1(a)). We found that the Fermi level shifts with the increase of Co dopant, in accord with the carrier doping behavior, similar to the case in ``122" type iron-pnictides. At heavily overdoped regime ($x=0.109$), the central hole pockets disappear and an electron pocket emerges around $\Gamma$. From the optimally-doped to the heavily overdoped non-superconducting phase, the overall band dispersions are barely affected, implying that the local interactions remain the same, though the superconductivity changes drastically.  Our data suggest that a balance between itinerancy of mobile carriers and local correlations may play an important role in the underlying physics of superconductivity.

The high-quality single crystals of NaFe$_{1-x}$Co$_x$As were grown using flux method with NaAs as the flux \cite{Wang}. The chemical composition of the single crystals was determined by energy dispersive X-ray spectroscopy with a standard instrument error $\sim$10$\%$. ARPES experiments were performed with \emph{s} photon polarization at the ARPES beamline of the National Synchrotron Radiation Laboratory at Hefei, China, using a Scienta R4000 electron spectrometer. The angular resolution was 0.3 degrees and the combined instrumental energy resolution was better than 20 meV. All samples were cleaved and measured at 25 K under a vacuum better than $5\times10^{-11}$ mbar. The data were taken within one hour after each cleave to eliminate the sample aging effects which could act in a fashion of hole doping.

\begin{figure}
\includegraphics[scale=0.75]{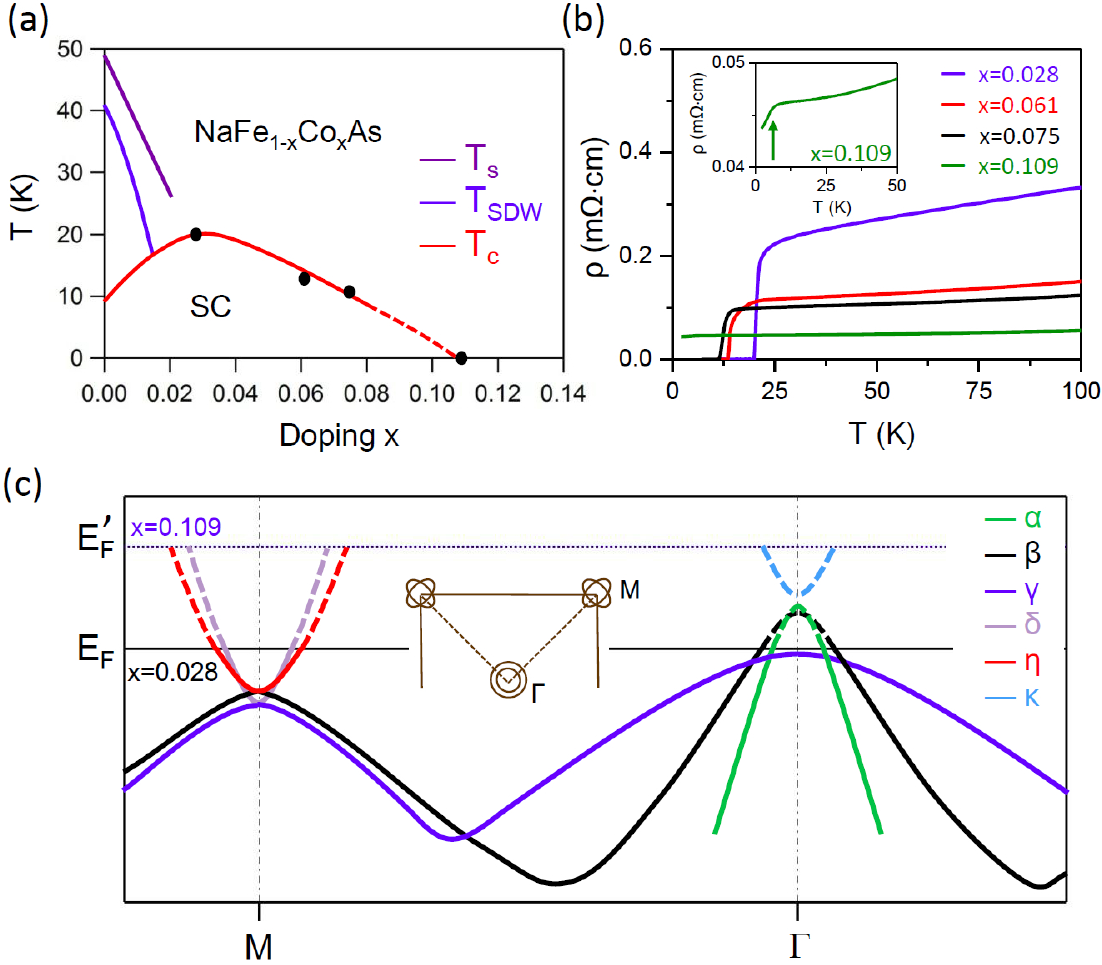}
\caption{(a) Phase diagram of NaFe$_{1-x}$Co$_x$As. The structural (T$_s$), spin-density-wave (T$_{SDW}$) and superconducting (T$_c$) transitions were determined by resistivity measurements \cite{Wang}. The T$_c$'s and compositions of our four compounds are marked by black dots. (b) In-plane resistivities of NaFe$_{1-x}$Co$_x$As. The inset shows a drop of resistivity at 6 K in the $x=0.109$ compound. (c) A schematic band structure of NaFe$_{1-x}$Co$_x$As. The solid lines depict the band structure in the $x=0.028$ compound, and the dashed lines describe the band structure that appears in $x=0.109$ samples. The inset sketches the Fermi surface of $x=0.028$ samples. }
\label{Fig1}
\end{figure}

Based on the transport measurements from the same NaFe$_{1-x}$Co$_x$As batches \cite{Wang}, we plot the electronic phase diagram in Fig. 1(a). The in-plane resistivities of the four compounds we studied are shown in Fig. 1(b). We note that there lacks data points in the overdoped region to determine the exact location where the T$_c$ drops to zero. In the inset of Fig. 1(b), we display the low-temperature data of the $x=0.109$ compound, which shows a decrease in resistivity below 6 K and hints a tendency toward superconductivity. We have measured several $x=0.109$ samples to verify this feature, and it was observed in each of them. Therefore, we argue that the $x=0.109$ compound is very close to the edge of the superconducting dome.

The solid lines in Fig. 1(c) show a sketch of band structure of NaFe$_{1-x}$Co$_x$As ($x=0.028$), which consists of three hole-like bands ($\alpha,\beta,\gamma$) around the zone center $\Gamma$ and two electron-like bands ($\delta,\eta$) around the zone corner M. As will be shown later, the Co dopant will shift the Fermi level significantly and an electron pocket ($\kappa$) will emerge around $\Gamma$ in $x=0.109$ samples. We plot dashed lines above the Fermi energy of the $x=0.028$ compound to describe the band structure in the heavily overdoped samples.

\begin{figure*}
\includegraphics[scale=0.95]{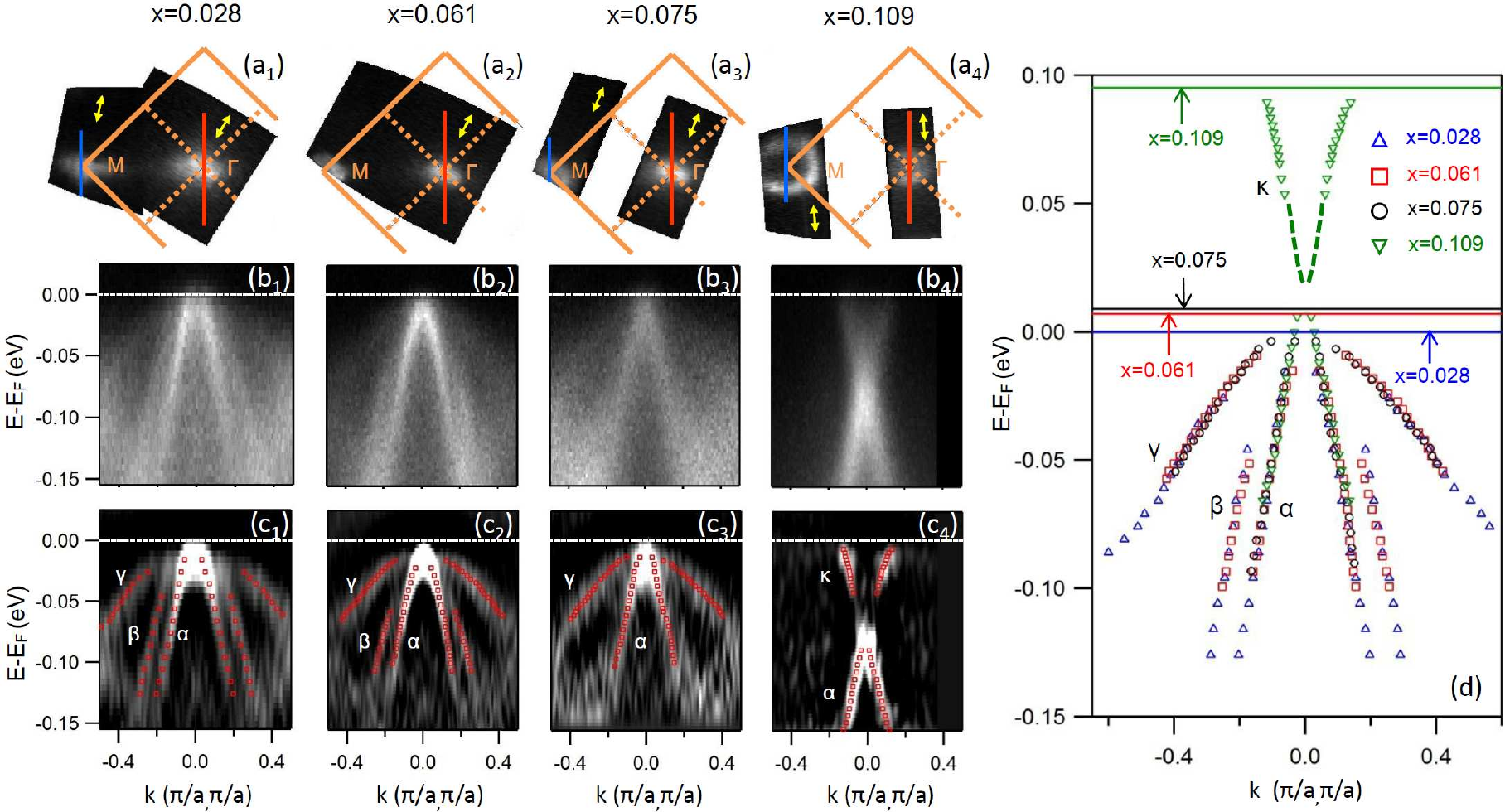}
\caption{(a$_1$-a$_4$) Fermi surface maps of NaFe$_{1-x}$Co$_x$As with various dopings. All data were taken at T = 25K with 20 eV photons of \emph{s} polarization. The ARPES intensity maps were obtained by integrating over an energy window of E$_F$$\pm$3 meV. The yellow arrows indicate the polarization directions of the incident beam for each Fermi surface patch. (b$_1$-b$_4$) Band dispersion maps along the red cuts in panels a$_1$-a$_4$. (c$_1$-c$_4$) The 2$^{nd}$ derivative images of experimental data in panels b$_1$-b$_4$. (d) Band dispersions determined by the 2$^{nd}$ derivative images as indicated by the red symbols in panels c$_1$-c$_4$. The data for different doping levels are overlaid by rigid-band shifting with their Fermi levels denoted individually. The band bottom of the $\kappa$ band cannot be determined accurately, we use a dashed curve to make an estimate.}
\label{Fig2}
\end{figure*}

Figs. 2(a$_1$-a$_4$) show the Fermi surface maps for different Co concentrations. All data were measured at T = 25 K to eliminate the change of band dispersions due to the opening of superconducting gaps. This choice enables us to compare the band dispersions of various NaFe$_{1-x}$Co$_x$As systematically. The data taken along $\Gamma$-M (red lines in Figs. 2(a$_1$-a$_4$)) are displayed in Figs. 2(b$_1$-b$_4$), and the 2$^{nd}$ derivative images of these data are shown in Figs. 2(c$_1$-c$_4$) to emphasize the band dispersions. In optimally doped samples ($x=0.028$), the three hole-like bands around $\Gamma$ are clearly visible. With increasing Co doping, the $\alpha$ band persists all the way up to the doping of $x=0.109$, while the $\beta$ and $\gamma$ bands appear invisible in Figs. 2(b$_4$, c$_4$). We argue that this difference could be an artificial effect arising from the photoemission matrix elements. In our measurements, the relative geometry between the photon polarization and orbital symmetries of Fe 3\emph{d} bands varies, which can highlight or weaken a band with certain orbital symmetries with respect to the light polarization \cite{ZhangPRB,Yi}.

One distinctive feature in the heavily overdoped non-superconducting compound ($x=0.109$) is the existence of an electron pocket ($\kappa$) at the zone center. Such an electron pocket has been observed in other heavily electron-doped iron pnictides \cite{LiuPRB,ZhangNatmaterials}. In NaFeAs, the $\kappa$ band is consistent with band structure calculations that show an electron-like band right above the hole-like bands around the zone center \cite{Kusakabe}. Here, we cannot clearly resolve the band bottom of the $\kappa$ band, so we use a dashed line to make an approximation in Fig. 2(d). Although the $\kappa$ band replaces the hole-like Fermi surface and breaks the quasiparticle scattering channel connecting the hole pockets around $\Gamma$ and electron pockets around M, we cannot come to the conclusion that it is responsible for the vanishing of superconductivity. The tendency toward superconductivity in the $x=0.109$ compound (see the inset of Fig. 1(b)) suggests that the superconductivity could survive at a doping level slightly below $x=0.109$, with a smaller $\kappa$ electron pocket at the zone center. Indeed, such a Fermi surface topology has been observed in A$_x$Fe$_{2-y}$Se$_2$ superconductors with an isotropic superconducting gap on the central $\kappa$ Fermi pocket \cite{ZhangNatmaterials,DMou,mXu,XPWang}.

Based on the 2$^{nd}$ derivative images shown in Figs. 2(c$_1$-c$_4$), we can determine the \emph{E} \emph{vs.} \emph{k} dispersions of $\alpha,\beta$ and $\gamma$ bands for various dopings as indicated by the red symbols. We note that there are uncertainties near the Fermi level where all three hole-like bands are very close to each other and cannot be well resolved in the 2$^{nd}$ derivative images. In Fig. 2(d), we plot all dispersions for various doping levels and overlay them by rigid-band shifting with their Fermi levels indicated respectively. It clearly shows that the $\gamma$ band dispersion remains almost the same over an energy scale of 50 - 100 meV except rigid shifting in energy, despite the fact that the compound changes from an optimally doped superconductor ($x=0.028$) to a heavily overdoped non-superconducting metal ($x=0.109$). The dispersions of $\alpha$ and $\beta$ bands exhibit the identical doping-independent behavior, though they cannot be determined by the $x=0.109$ data. However, as will be discussed later, it is highly plausible that the $\alpha$ and $\beta$ band dispersions will follow the doping-independent behavior of the $\gamma$ band in the $x=0.109$ compound,

From $x=0.028$ to $x=0.061, 0.075$, the Fermi level shifts by 7$\pm3$ meV and 9$\pm3$ meV respectively, and it increases by 95$\pm3$ meV in the $x=0.109$ compound, suggesting that extra electrons are released by Co dopant. This trend has been observed on samples from the same batches by scanning tunneling microscopy, which shows that the loci of the minimal value and a hump on the negative bias side in the dI/dV curves shift by $\sim$ 100 meV from $x=0.028$ to $x=0.109$ \cite{xZhou}. Such a Fermi level shifting can also be found in the electron-like bands near the zone corner M. In Fig. 3, we show the data along the blue cuts in Figs. 2(a$_1$, a$_3$, a$_4$). Compared to $x=0.028, 0.075$ compounds, the Fermi level moves up by $\sim$ 100 meV in $x=0.109$ samples, though the bands cannot be clearly resolved at low doping levels. There are two electron-like bands around M in band calculations, while they cannot be distinguished in Fig. 3(c), suggesting they are nearly degenerated.

\begin{figure}
\includegraphics[scale=0.75]{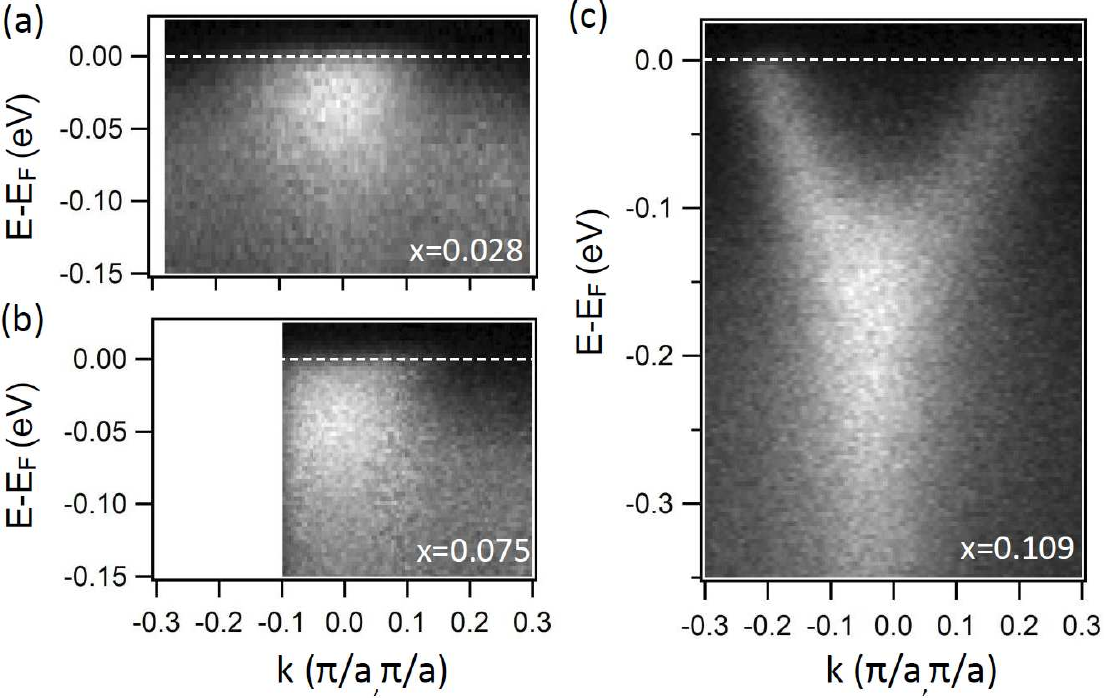}
\caption{(a-c) Band dispersion maps around the zone corner M for $x=0.028, 0.075, 0.109$, respectively, taken along the blue cuts in Figs. 2(a$_1$,a$_3$,a$_4$).}
\label{Fig3}
\end{figure}

The doping-independent behavior of band renormalizations in Fig. 2(d) provides some hints on the local interactions. Compared with LDA band calculations, the near E$_F$ bands in iron pnictides are usually found to be renormalized by a factor of $\sim$ 2-3 \cite{DHLu}, indicating notable electron correlations. Generally speaking, the local interactions in real space result in effects throughout the momentum space, which can renormalize the overall bandwidth. In iron pnictides, the most relevant sources for local interactions are the Hubbard \emph{U} and Hund's coupling \emph{J}. Recent X-ray absorption and resonant inelastic X-ray scattering experiments on BaFe$_2$As$_2$ gave \emph{U} $\lesssim$ 2 eV and \emph{J} $\approx$ 0.8 eV \cite{WLYang}, which are comparable with the bandwidths of the Fe 3\emph{d} bands near E$_F$. It is very reasonable to believe that a similar situation exists in NaFeAs, and the overall band dispersions are largely renormalized by the \emph{U} and \emph{J}. In spite of the change of superconducting properties with carrier doping, as shown in Fig. 2(d), the hole-like band dispersions remain basically the same over an energy scale of 50-100 meV. Potentially this behavior could occur throughout the momentum space within a significant energy scale, because the local correlations in real space correspond to a global variation in momentum space. Our data indicate that the local electronic and/or magnetic interactions barely vary with carrier doping. This finding is a good starting point, from where we can examine the connections between superconductivity and other degrees of freedom.

When look into novel properties of correlated electron systems, one has to consider both the local correlations and itinerancy of mobile carriers. It has been argued that a right balance between two of them is a key to a high T$_c$ \cite{DNBasov}. In NaFe$_{1-x}$Co$_x$As ($x=0.028 - 0.109$), since the local interactions are independent of carrier doping, the variations of itinerancy are likely responsible for the decline of T$_c$. As we have mentioned, similar to A$_x$Fe$_{2-y}$Se$_2$ superconductors, the central $\kappa$ pocket of the $x=0.109$ compound may not certainly lead to the destruction of superconductivity, though it departs from the theoretical scenarios involving hole pockets around the zone center. However, its effects need to be evaluated and further investigations are required to address this issue clearly. On the other hand, the Fermi velocity is a crucial parameter of itinerant properties. In $x=0.028 - 0.075$ samples, the Fermi energy is close to the bottoms of electron-like bands at the zone corner M (see Figs. 3(a,b)), and thus the Fermi velocities possess relatively low values. Since the band dispersions cannot be clearly resolved in Figs. 3(a,b), we use the data in ref.\cite{zhliu} to make an estimate, which gives a upper limit of  0.4-0.5 eV$\cdot${\AA} for the Fermi velocity. This value is similar to those in (Tl$_{0.58}$Rb$_{0.42}$)Fe$_{1.72}$Se$_2$ \cite{DMou}, K$_{0.8}$Fe$_{1.7}$Se$_2$ \cite{TQian} and single-layer FeSe \cite{DLiu}, where the electron pockets around the zone corner is the most relevant for superconductivity. In $x=0.109$ samples, the large E$_F$ shifting by carrier doping results in a significantly increase in the Fermi velocity of 50\% - 100\% (Fig. 3(c)). This change of Fermi velocities could break the balance between local correlations and carrier itinerancy and suppress T$_c$.

 Althought we have stressed the local interactions and itinerant properties of electrons, there are other ingredients that could play important roles. For example, antiferromagnetic fluctuations are considered to be a key factor of superconductivity in many theoretical models, and it has been suggested that they cause the linear temperature dependence of magnetic susceptibility in iron pnictides \cite{GMZhang}. In the $x=0.109$ crystals, the susceptibility deviates from the linear temperature behavior and may imply the absence of antiferromagnetic fluctuations \cite{Wang}. This helps to explain the disappearance of superconductivity. Moreover, the Co doping could introduce disorders that possibly impair T$_c$ in the heavily overdoped region, though it has been shown that this effect can be ignored in some iron pnictides \cite{Katase}. More studies are required to clarify various issues here.

In summary, we have studied the evolution of band structure of NaFe$_{1-x}$Co$_x$As from the optimal doping ($x=0.028$) to the heavily overdoped non-superconducting regime ($x=0.109$). In this doping range, there are no impacts of the structural transition or long-range magnetically ordered phase, and the dominant effects arise from carrier doping induced by Co substitution. With increasing carriers, the Fermi level shifts and the Fermi surface changes drastically. The hole-like bands around $\Gamma$ move down to deeper binding energies, and an electron pocket appears instead at the zone center in the $x=0.109$ compound. The sizes of electron pockets around the zone corner increase with Co doping. The consequent Fermi surface in the heavily overdoped regime consists of electron pockets only and are drastically different from the band structure in low-doping regime. Regardless the Fermi energy shifting, the overall band dispersions remain almost the same, suggesting that the local electronic or magnetic correlations are essentially independent of carrier doping in the range we studied. Accompanying the decrease of T$_c$ with increasing Co doping, the dominant changes are the Fermi surface topology and Fermi velocities, suggesting that the itinerant properties of mobile carriers and their balance with local correlations could be of importance for the superconductivity in iron pnictides.

G. B. Z and Z. S acknowledge D. L. Feng and B. P. Xie for their tremendous support during the construction of the ARPES system at the National Synchrotron Radiation Laboratory. Z. S is grateful to J. P. Hu for enlightening discussions and his critical comments. This work was supported by National Natural Science Foundation of China (Grant No. 11190022, 11174264, 11190021), the National Basic Research Program of China (973 Program, Grant No. 2012CB922002, 2012CB922004), and the Chinese Academy of Sciences. Z. S acknowledges the support by the Fundamental Research Funds for the Central Universities (Grant No. WK 2310000024).


\begin{thebibliography}{}

\bibitem{Kamihara}Y. Kamihara \emph{et al.}, J. Am. Chem. Soc. \textbf{130}, 3296 (2008).
\bibitem{xhchen}X. H. Chen \emph{et al.}, Nature \textbf{453},761-762 (2008).
\bibitem{Rotter}M. Rotter  \emph{et al.},  Phys. Rev. Lett. \textbf{101}, 107006 (2008).
\bibitem{Canfield}P. C. Canfield \emph{et al.}, Phys. Rev. B \textbf{80}, 060501(R) (2009).
\bibitem{Shuai}J. Shuai \emph{et al.}, J. Phys.: Condens. Matter \textbf{21}, 382203 (2009).
\bibitem{Schnelle}W. Schnelle \emph{et al.}, Phys. Rev. B \textbf{79}, 214516 (2009).
\bibitem{LiuNat}C. Liu \emph{et al.}, Nat. Phys. \textbf{6}, 419 (2010).
\bibitem{LiuPRB}C. Liu \emph{et al.}, Phys. Rev. B \textbf{84}, 020509 (2011).
\bibitem{Dhaka}R. S. Dhaka \emph{et al.}, Phys. Rev. Lett. \textbf{107}, 267002 (2011).
\bibitem{Xu}N. Xu \emph{et al.}, arxiv:1203.4699.
\bibitem{ZhangNat}Y. Zhang \emph{et al.}, Nature Physics \textbf{8}, 371 (2012).
\bibitem{Yoshida}T. Yoshida \emph{et al.}, Phys. Rev. Lett. \textbf{106}, 117001 (2011).
\bibitem{Thirupathaiah}S. Thirupathaiah \emph{et al.}, Phys. Rev. B \textbf{84}, 014531 (2011).
\bibitem{Neupane}M. Neupane \emph{et al.}, Phys. Rev. B \textbf{83}, 094522 (2011).
\bibitem{Ideta}S. Ideta \emph{et al.}, arxiv:1205.1889.
\bibitem{Wadati}H. Wadati \emph{et al.}, Phys. Rev. Lett. \textbf{105}, 157004 (2010).
\bibitem{Parker}D. R. Parker \emph{et al.}, Phys. Rev. Lett. \textbf{104}, 057007 (2010).
\bibitem{zhliu}Z. H. Liu \emph{et al.}, Phys. Rev. B \textbf{84}, 064519 (2011).
\bibitem{Wright}J. D. Wright \emph{et al.}, Phys. Rev. B \textbf{85}, 054503 (2012).
\bibitem{syZhou}S. Y. Zhou \emph{et al.}, arxiv:1204.3440.
\bibitem{Wang}A. F. Wang \emph{et al.}, Phys. Rev. B \textbf{85}, 224521 (2012).
\bibitem{xZhou}X. Zhou \emph{et al.}, arxiv:1204.4237.
\bibitem{ZhangPRB}Y. Zhang \emph{et al.}, Phys. Rev. B \textbf{85}, 085121 (2012).
\bibitem{Yi}M. Yi \emph{et al.}, arxiv:1111.6134.
\bibitem{ZhangNatmaterials}Y. Zhang \emph{et al.}, Nature Materials \textbf{10}, 273-277 (2011).
\bibitem{Kusakabe}K. Kusakabe and A. Nakanishi, J. Phys. Soc. Jpn. \textbf{78}, 124712 (2009).
\bibitem{DMou}D. X. Mou, L. Zhao, X. J. Zhou, Front. Phys. \textbf{6}, 410 (2011).
\bibitem{mXu}M. Xu \emph{et al.}, arxiv:1205.0787.
\bibitem{XPWang}X. -P. Wang \emph{et al.}, arxiv:1205.0996.
\bibitem{DHLu}D. H. Lu \emph{et al.}, Nature \textbf{455}, 81 (2008).
\bibitem{WLYang}W. L. Yang \emph{et al.}, Phys. Rev. B \textbf{80}, 014508 (2009).
\bibitem{DNBasov}D. N. Basov and A. V. Chubukov, Nature Physics \textbf{7},272 (2011).
\bibitem{TQian}T. Qian \emph{et al.}, Phys. Rev. Lett. \textbf{106}, 187001 (2011).
\bibitem{DLiu}D. Liu \emph{et al.}, arxiv:1202.5849.
\bibitem{GMZhang}G. M. Zhang \emph{et al.}, Europhys. Lett. \textbf{86}, 37006 (2009).
\bibitem{Katase}T. Katase \emph{et al.}, Phys. Rev. B \textbf{85}, 140516 (2012).

\end{thebibliography}
\end{document}